\documentclass[epsfig]{revtex4}
\usepackage{amsmath}
\usepackage{graphicx}
\begin{document}

\title{Multiparty hierarchical quantum-information splitting}
\author{Xin-Wen Wang,
 Deng-Yu Zhang, Shi-Qing Tang, and Li-Jun Xie}
\address{Department of Physics and Electronic Information Science,
Hengyang Normal University, Hengyang 421008, China}

\begin{abstract}
We propose a scheme for multiparty hierarchical quantum-information
splitting (QIS) with a multipartite entangled state, where a boss
distributes a secret quantum state to two grades of agents
asymmetrically. The agents who belong to different grades have
different authorities for recovering boss's secret. Except for
boss's Bell-state measurement, no nonlocal operation is involved.
The presented scheme is also shown to be secure against
eavesdropping. Such a hierarchical QIS is expected to find useful
applications in the field of modern multipartite quantum
cryptography.
 \\


\end{abstract}

\maketitle

A fundamental ingredient for implementation of quantum technologies
is the ability to faithfully transmit quantum states among quantum
mechanical systems which are even far apart. Quantum-information
splitting (QIS, also be referred to as quantum-secret sharing or
quantum-state sharing in the literature), first introduced by
Hillery, Bu\v{z}ek, and Berthiaume (HBB) \cite{59PRA1829}, is a
typical way for quantum state transfer, in which a secret quantum
state is distributed by quantum teleportation \cite{70PRL1895} from
a boss to more than one agents so that any one of them can recover
the state with assistance of the others. QIS is a generalization of
classical-secret sharing to quantum scenario. Classical-secret
sharing is one of the most important information-theoretically
secure cryptographic protocols and is germane to online auctions,
electronic voting, shared electronic banking, cooperative activation
of bombs, and so on. Also, QIS has extensive applications in
quantum-information science, such as creating joint checking
accounts containing quantum money \cite{quantum money}, secure
distributed quantum computation \cite{0801.1544,0906.2297}, and so
on.

 In the original HBB QIS proposal with the quantum channel
being a three-qubit Greenberger-Horne-Zeilinger (GHZ) state
\cite{GHZ}, the collaboration of two agents is implemented by means
of classical communication about their single-particle measurement
outcomes. This idea can be directly generalized to the case of $N$
agents by using an $(N+1)$-particle GHZ state, or by the way of
Ref.~\cite{59PRA162}. These schemes are ($N$, $N$)-threshold schemes
where all the $N$ agents need collaborating in order to recover the
secret state. Soon after, Cleve, Gottesman, and Lo (CGL)
\cite{83PRL648} proposed another type of QIS scheme with the idea of
quantum error-correcting codes. The CGL scheme is a ($K$,
$N$)-threshold scheme where $K$ ($[N/2]<K\leq N$) of $N$ agents can
extract the quantum information of the original secret state by
cooperation. The CGL QIS scheme, however, needs the cooperated
agents to make nonlocal operations on their particles. That is, the
$K$ cooperated agents need to transmit their $K$ particles to one
laboratory and perform a collective operation (decoding operation)
on them. In the last decade, both of the above two QIS ideas have
triggered significant research activity (see, e.g.,
\cite{61PRA042311,64PRA042311,71PRA012328,81PRA052333,74PRA054303,324PLA420,354PLA190,39JPB1975,41JPB015503,42JPA115303,8QIP431}),
and some schemes have already been experimentally realized
\cite{430N54,92PRL177903}.

 A more general QIS scheme should involve the asymmetry between
the powers of the different participants, where only particular
subsets of the participants can recover the secret quantum state by
cooperation. For instance, one might consider a scheme with three
participants $A$, $B$, and $C$. The sets ($A$, $B$) or ($A$, $C$)
can reconstruct the secret state, but ($B$, $C$) can not. In this
example, the presence of $A$ is essential to reconstructing the
secret state, but not sufficient. The asymmetric QIS has been
intensively studied with the idea of quantum error-correcting codes
based on multi-particle nonlocal operations
\cite{61PRA042311,64PRA042311,71PRA012328,81PRA052333}. Although an
asymmetric three-party QIS scheme has also been put forward
\cite{283OC1196} with a four-partite entangled state
\cite{96PRL060502,78PRA024301,282OC1052} and classical
communications, there has been no intensive study on this issue.

In this paper, we propose a multiparty asymmetric QIS scheme with a
multipartite-entanglement channel and classical communications. The
scheme involves two grades of agents $G_1=\{$Bob$_1$, Bob$_2$,
$\cdots$, Bob$_m\}$ and $G_2=\{$Charlie$_1$, Charlie$_2$, $\cdots$,
Charlie$_n\}$. For getting boss's (Alice's) secret state, one of
Bobs needs the collaboration of the other Bobs and any one of
Charlies, while one of Charlies needs the collaboration of all the
other $m+n-1$ agents. This indicates that the agents who belong to
different grades have different authorities to recover boss's secret
state. Such a type of QIS is referred to as the hierarchical QIS
hereafter. Note that the collaboration of the agents is based on
single-particle measurements and classical communications, and they
do not need to make any nonlocal operation.

We now introduce the hierarchical QIS in detail. The quantum channel
shared among Alice, Bobs, and Charlies is an $(1+m+n)$-qubit graph
state given by
\begin{eqnarray}
\label{QC}
|\mathcal{G}\rangle_{1+m+n}
  &=&\frac{1}{2}(|0_A0_{B_1}0_{B_2}\cdots 0_{B_m}0_{C_1}0_{C_2}\cdots 0_{C_n}\rangle\nonumber\\
 &&+|0_A0_{B_1}0_{B_2}\cdots 0_{B_m}1_{C_1}1_{C_2}\cdots 1_{C_n}\rangle\nonumber\\
&&+|1_A1_{B_1}1_{B_2}\cdots 1_{B_m}0_{C_1}0_{C_2}\cdots 0_{C_n}\rangle\nonumber\\
&&   -|1_A1_{B_1}1_{B_2}\cdots 1_{B_m}1_{C_1}1_{C_2}\cdots 1_{C_n}\rangle).
\end{eqnarray}
By performing, respectively, each qubit except $B_1$ and $C_1$ a
Hardamard transformation
$H=(|0\rangle\langle0|+|1\rangle\langle0|+|0\rangle\langle1|-|1\rangle\langle1|)/\sqrt{2}$,
one can transform the graph state of Eq.~(\ref{QC}) into the
standard form \cite{69PRA062311}
\begin{eqnarray}
 |stand.\rangle &=&\frac{1}{2^{(1+m+n)/2}}(|0_A\rangle+|1_A\rangle\sigma^z_{B_1})\nonumber\\
 && \otimes (|0_{B_1}\rangle+|1_{B_1}\rangle\sigma^z_{B_2}\cdots \sigma^z_{B_m}\sigma^z_{C_1})\nonumber\\
 && \otimes (|0_{B_2}\rangle+|1_{B_2}\rangle)\cdots (|0_{B_m}\rangle+|1_{B_m}\rangle)\nonumber\\
 && \otimes (|0_{C_1}\rangle+|1_{C_1}\rangle\sigma^z_{C_2}\cdots \sigma^z_{C_n})\nonumber\\
&& \otimes (|0_{C_2}\rangle+|1_{C_2}\rangle)\cdots (|0_{C_n}\rangle+|1_{C_n}\rangle),
\end{eqnarray}
where $\sigma^z_{j}=|0_j\rangle\langle 0_j|-|1_j\rangle\langle 1_j|$
is the usual Pauli operator. Here qubit A is held by Alice, qubit
$B_i$ by Bob$_i$ ($i=1,2,\cdots,m$), and qubit $C_{i'}$ by
Charlie$_{i'}$ ($i'=1,2,\cdots,n$). The quantum state to be
distributed is described by
\begin{equation}
  |\xi\rangle_S=\alpha|0_S\rangle+\beta|1_S\rangle,~~~~|\alpha|^2+|\beta|^2=1.
\end{equation}
The state of the whole system is
\begin{eqnarray}
\label{whole}
  |\mathcal{W}\rangle &=&|\xi\rangle_S\otimes|\mathcal{G}\rangle_{1+m+n}\nonumber\\
  &=&\frac{1}{\sqrt{2}}(\alpha|0_S0_A\rangle|\varphi\rangle^0_{m+n}+\alpha|0_S1_A\rangle|\varphi\rangle^1_{m+n}\nonumber\\
  &&+\beta|1_S0_A\rangle|\varphi\rangle^0_{m+n}+\beta|1_S1_A\rangle|\varphi\rangle^1_{m+n}),
\end{eqnarray}
where
\begin{eqnarray}
|\varphi\rangle^0_{m+n}&=&\frac{1}{\sqrt{2}}|0_{B_1}0_{B_2}\cdots 0_{B_m}
   \rangle(|0_{C_1}0_{C_2}\cdots 0_{C_n}\rangle\nonumber\\
   &&+|1_{C_1}1_{C_2}\cdots 1_{C_n}\rangle),\nonumber\\
|\varphi\rangle^1_{m+n}&=&\frac{1}{\sqrt{2}}|1_{B_1}1_{B_2}\cdots 1_{B_m}
   \rangle(|0_{C_1}0_{C_2}\cdots 0_{C_n}\rangle\nonumber\\
   &&-|1_{C_1}1_{C_2}\cdots 1_{C_n}\rangle).
\end{eqnarray}
In order to implement QIS, Alice performs a joint measurement on her
two qubits $S$ and $A$ in the Bell basis
$\{|\Phi\rangle^{\pm}_{SA},|\Psi\rangle^{\pm}_{SA}\}$, and then
informs agents of the outcome by classical communication. The four
Bell states are given by
\begin{eqnarray}
  |\Phi\rangle^{\pm}_{SA}=\frac{1}{\sqrt{2}}(|0_S0_A\rangle \pm |1_S1_A\rangle),\nonumber\\
  |\Psi\rangle^{\pm}_{SA}=\frac{1}{\sqrt{2}}(|0_S1_A\rangle \pm |1_S0_A\rangle).
\end{eqnarray}
For Alice's four possible measurement outcomes,
$|\Phi\rangle^{\pm}_{SA}$ or $|\Psi\rangle^{\pm}_{SA}$, the qubits
held by Bobs and Charlies collapse correspondingly into the
following entangled states:
\begin{eqnarray}
  |\phi\rangle^{\pm}_{m+n}=\alpha|\varphi\rangle^0_{m+n}\pm \beta|\varphi\rangle^1_{m+n}, \nonumber\\
 |\psi\rangle^{\pm}_{m+n}=\alpha|\varphi\rangle^1_{m+n}\pm \beta|\varphi\rangle^0_{m+n}.
\end{eqnarray}
The non-cloning theorem \cite{239N802,92PLA271} allows only one
qubit to be in the secret state $|\xi\rangle$, so that any one of
the $m+n$ agents, but not all, can recover such a state.

First, we assume that they agree to let Bob$_1$ possess the secret.
We rewrite $|\phi\rangle^{\pm}_{m+n}$ and $|\psi\rangle^{\pm}_{m+n}$
as
\begin{eqnarray}
\label{Bob1}
|\phi\rangle^{\pm}_{m+n}&=&\frac{1}{2}\left[(\alpha|0_{B_1}\rangle\pm\beta|1_{B_1}\rangle)
  \left(|m-1,m-1\rangle_{B_2\cdots B_m}|0_{C_1}\cdots 0_{C_n}\rangle\right.\right.\nonumber\\
  &&\left.+|m-2,m-1\rangle_{B_2\cdots B_m}|1_{C_1}\cdots 1_{C_n}\rangle\right)\nonumber\\
&&+(\alpha|0_{B_1}\rangle\mp\beta|1_{B_1}\rangle)
  \left(|m-1,m-1\rangle_{B_2\cdots B_m}|1_{C_1}\cdots 1_{C_n}\rangle \right.\nonumber\\
  &&\left.\left.+|m-2,m-1\rangle_{B_2\cdots B_m}|0_{C_1}\cdots 0_{C_n}\rangle\right)\right],\nonumber\\
|\psi\rangle^{\pm}_{m+n}&=&\frac{1}{2}\left[(\alpha|1_{B_1}\rangle\pm\beta|0_{B_1}\rangle)
  \left(m-1,m-1\rangle_{B_2\cdots B_m}|0_{C_1}\cdots 0_{C_n}\rangle\right.\right.\nonumber\\
  &&\left.+|m-2,m-1\rangle_{B_2\cdots B_m}|1_{C_1}\cdots 1_{C_n}\rangle\right)\nonumber\\
&&+(\alpha|1_{B_1}\rangle\mp\beta|0_{B_1}\rangle)
  \left(|m-1,m-1\rangle_{B_2\cdots B_m}|1_{C_1}\cdots 1_{C_n}\rangle\right.\nonumber\\
 && \left.\left.+|m-2,m-1\rangle_{B_2\cdots B_m}|0_{C_1}\cdots 0_{C_n}\rangle\right)\right],
\end{eqnarray}
where $|\pm\rangle=(|0\rangle\pm |1\rangle)/\sqrt{2}$ and the
following notations are used:
\begin{eqnarray}
&&|X-1,X\rangle=\frac{1}{2^{(X-1)/2}}\sum_{k=0}^{[(X-1)/2]}\sqrt{C_{X}^{2k+1}}|\{-,2k+1\};\{+,X-2k-1\}\rangle,\nonumber\\
&&|X,X\rangle=\frac{1}{2^{(X-1)/2}}\sum_{k=0}^{[X/2]}\sqrt{C_{X}^{2k}}|\{-,2k\};\{+,X-2k\}\rangle,
\end{eqnarray}
with $X$ being an integer, $[x/2]$ ($x=X,X-1$) being the integer
part of $x/2$, $C_X^K=X!/K!(X-K)!$ ($K=2k,2k+1$) being the
combinational coefficient, and $|\{-,K\};\{+,X-K\}\rangle$ denoting
all the totally symmetric states including $K$ ``$-$'' and $X-K$
``$+$'' (e.g.,
$|\{-,1\};\{+,2-1\}\rangle=(|-+\rangle+|+-\rangle)/\sqrt{2}$).
 It can be seen from Eq.~(\ref{Bob1}) that the other $m-1$ Bobs and
one of Charlies (denoted by Charlie$^*$) can assist Bob$_1$ in
recovering Alice's secret state. Particularly, Bobs measure their
qubits in the basis $\{|+\rangle,|-\rangle\}$ and Charlie$^*$
measures his qubit in the basis $\{|0\rangle,|1\rangle\}$, and
inform Bob$_1$ of their measurement outcomes. For reconstructing the
state $|\xi\rangle$, Bob$_1$ needs to perform one of the unitary
transformations $\{I,\sigma^x, i\sigma^y,\sigma^z\}$ on qubit $B_1$,
where $I$ is the identity operator and $\sigma^{x,y,z}$ are the
usual Pauli operators. The one-to-one correspondence between
Bob$_1$'s operations and the measurement outcomes of Alice,
Charlie$^*$, and the other Bobs is shown in Table 1. $V_{sum}=
V_{G_1}\oplus V_{Charlie^*}$ with $\oplus$ being the modulo-2-sum,
where $V_{Charlie^*}$ and $V_{G_1}$ denote the values of the
outcomes obtained by Charlie$^*$ and Bobs, respectively. The states
$\{|\{-,2k\};\{+,X-2k\}\rangle,|0\rangle\}$ are encoded as the value
``0'' and $\{|\{-,2k+1\};\{+,X-2k-1\}\rangle,|1\rangle\}$ as ``1''.
For instance, Alice's Bell-state measurement outcome is
$|\Phi^+\rangle_{SA}$, even number of Bobs get the outcome
$|-\rangle$ ($V_{G_1}=0$), and Charlie$^*$ get $|1\rangle$
($V_{Charlie^*}=1$), then Bob$_1$ needs to make $\sigma^z$ operation
on qubit $B_1$ for reconstructing Alice's secret state $|\xi\rangle$
on it. The above results are also applicable to the case where one
of the other Bobs is deputed to possess Alice's secret because
$|\mathcal{G}\rangle_{1+m+n}$ is unchanged under the permutation of
qubits $\{B_1,B_2,\cdots, B_m\}$, which indicates that all the $m$
Bobs have the same status in the QIS protocol.

\begin{table}
\tabcolsep 5mm \caption{\label{tab:1}The one-to-one correspondence
between Bob$_1$'s unitary operations and the measurement outcomes of
Alice, Charlie$^*$, and the other Bobs.}
\begin{center}
\begin{tabular}{lll}
\hline Alice's outcomes & $V_{sum}$ & Operations\\
 \hline
  $|\Phi\rangle^+_{SA} ~(|\Phi\rangle^-_{SA})$ & 0~(1) & $I$    \\
   $|\Phi\rangle^+_{SA}~ (|\Phi\rangle^-_{SA})$ & 1~(0) & $\sigma^z$   \\
   $|\Psi\rangle^+_{SA}~ (|\Psi\rangle^-_{SA})$ & 0~(1) & $\sigma^x$    \\
$|\Psi\rangle^+_{SA} ~(|\Psi\rangle^-_{SA})$ & 1~(0) & $i\sigma^y$  \\
\hline
\end{tabular}
\end{center}
\end{table}

Now, we consider the case that they agree to let Charlie$_1$ recover
Alice's secret state. Then the states $|\phi^{\pm}\rangle_{m+n}$ and
$|\psi^{\pm}\rangle_{m+n}$ can be rewritten as
\begin{eqnarray}
\label{Charlie1}
|\phi^{\pm}\rangle_{m+n}&=&\frac{1}{2}\left[(\alpha|+_{C_1}\rangle\pm\beta|-_{C_1}\rangle)
  |m,m\rangle_{B_1\cdots B_m}|n-1,n-1\rangle_{C_2\cdots C_n}\right.\nonumber\\
&&+(\alpha|-_{C_1}\rangle\pm\beta|+_{C_1}\rangle)|m,m\rangle_{B_1\cdots B_m}|n-2,n-1\rangle_{C_2\cdots C_n}\nonumber\\
&&+(\alpha|+_{C_1}\rangle\mp\beta|-_{C_1}\rangle)|m-1,m\rangle_{B_1\cdots B_m}
    |n-1,n-1\rangle_{C_2\cdots C_n}\nonumber\\
&&\left.+(\alpha|-_{C_1}\rangle\mp\beta|+_{C_1}\rangle)
  |m-1,m\rangle_{B_1\cdots B_m}|n-2,n-1\rangle_{C_2\cdots C_n}\right],\nonumber\\
|\psi^{\pm}\rangle_{m+n}&=&\frac{1}{2}\left[(\alpha|-_{C_1}\rangle\pm\beta|+_{C_1}\rangle)
   |m,m\rangle_{B_1\cdots B_m}|n-1,n-1\rangle_{C_2\cdots C_n}\right.\nonumber\\
&&+(\alpha|+_{C_1}\rangle\pm\beta|-_{C_1}\rangle)|m,m\rangle_{B_1\cdots B_m}|n-2,n-1\rangle_{C_2\cdots C_n}\nonumber\\
&&-(\alpha|-_{C_1}\rangle\mp\beta|+_{C_1}\rangle)|m-1,m\rangle_{B_1\cdots B_m}
   |n-1,n-1\rangle_{C_2\cdots C_n}\nonumber\\
&&\left.-(\alpha|+_{C_1}\rangle\mp\beta|-_{C_1}\rangle)
  |m-1,m\rangle_{B_1\cdots B_m}|n-2,n-1\rangle_{C_2\cdots C_n}\right].
\end{eqnarray}
It can be seen that Charlie$_1$ can reconstruct the state
$|\xi\rangle$ if and only if $m$ Bobs and the other $n-1$ Charlies
measure their qubits in the basis $\{|+\rangle,|-\rangle\}$ and
broadcast their outcomes. In other words, Charlie$_1$ needs the help
of all of the other agents for recovering Alice's secret. For
reconstructing the state $|\xi\rangle$, Charlie$_1$ needs to perform
one of the unitary transformations $\{H,\sigma^xH,
i\sigma^yH,\sigma^zH\}$ on qubit $C_1$. The one-to-one
correspondence between Charlie$_1$'s operations and the measurement
outcomes of Alice and the other agents is shown in Table 2.
$V_{G_1}$ and $V_{G_2}$ denote the values of the outcomes obtained
by Bobs and Charlies, respectively. The above results are also
applicable to the case where one of the other Charlies is deputed to
possess Alice's secret because $|\mathcal{G}\rangle_{1+m+n}$ is
unchanged under the permutation of qubits $\{C_1,C_2,\cdots, C_n\}$,
which indicates that all the $n$ Charlies have the same status in
the QIS protocol.

\begin{table}
\tabcolsep 5mm \caption{\label{tab:2}The one-to-one correspondence
between Charlie$_1$'s unitary operations and the measurement
outcomes of Alice, Bobs, and the other Charlies.}
\begin{center}
\begin{tabular}{llll}
\hline
 Alice's outcomes & $V_{G_1}$ & $V_{G_2}$ & Operations\\
 \hline
  $|\Phi\rangle^+_{SA} ~(|\Phi\rangle^-_{SA})$ & 0~(1) & 0 & $H$\\
   $|\Phi\rangle^+_{SA}~ (|\Phi\rangle^-_{SA})$ & 1~(0) &0 & $\sigma^zH$\\
    $|\Phi\rangle^+_{SA} ~(|\Phi\rangle^-_{SA})$ & 0~(1) & 1 & $\sigma^xH$\\
   $|\Phi\rangle^+_{SA}~ (|\Phi\rangle^-_{SA})$ & 1~(0) &1 & $i\sigma^yH$\\
   $|\Psi\rangle^+_{SA}~ (|\Psi\rangle^-_{SA})$ & 0~(1) &1& $H$  \cr
   $|\Psi\rangle^+_{SA} ~(|\Psi\rangle^-_{SA})$ & 1~(0)& 1 & $\sigma^zH$ \\
   $|\Psi\rangle^+_{SA}~ (|\Psi\rangle^-_{SA})$ & 0~(1) &0& $\sigma^xH$  \\
$|\Psi\rangle^+_{SA} ~(|\Psi\rangle^-_{SA})$ & 1~(0)& 0 &
$i\sigma^yH$ \\
\hline
\end{tabular}
\end{center}
\end{table}

According to former analysis, for recovering the secret state
$|\xi\rangle$, one of Bobs only needs the assistance of any one of
Charlies with the other Bobs, while one of Charlies needs the help
of all of Bobs with the other Charlies. Thus, their authorities for
getting Alice's secret are hierarchized, and Bobs are in a higher
position relative to Charlies. This result may be understood
partially from the picture as follows. After Alice's Bell-state
measurement, with outcomes $|\Phi^{\pm}\rangle_{SA}$ or
$|\Psi^{\pm}\rangle_{SA}$, Bobs' and Charlies' single-qubit
state-density matrices are, respectively,
\begin{eqnarray}
&&\rho^{|\Phi^{\pm}\rangle}_{Bob}=|\alpha|^2|0\rangle\langle 0|+|\beta|^2|1\rangle\langle 1|,\nonumber\\
&&\rho^{|\Psi^{\pm}\rangle}_{Bob}=|\beta|^2|0\rangle\langle 0|+|\alpha|^2|1\rangle\langle 1|,\nonumber\\
&&\rho_{Charlie}=\frac{1}{2}(|0\rangle\langle 0|+|1\rangle\langle 1|),
\end{eqnarray}
where the superscripts $|\Phi^{\pm}\rangle$ and $|\Psi^{\pm}\rangle$
denote Alice's measurement outcomes. It can be seen that each
Charlie knows nothing about the information of Alice's secret state
$|\xi\rangle$ without the collaboration of the other agents; each
Bob, however, has the amplitude information of $|\xi\rangle$ as long
as receiving Alice's Bell-state measurement outcome. This implies
that Alice's secret quantum state is distributed to Bobs and
Charlies asymmetrically. Naturally, the more information is known,
the less collaborations are needed.

We now give a brief discussion on the security of this scheme
against a potential eavesdropper (say Eve). Because of the
no-cloning theorem \cite{239N802,92PLA271} and entanglement monogamy
\cite{61PRA052306}, the only way for Eve to eavesdrop the secret
state is to take the intercept-resend attack. Particularly, Eve
intercepts all the qubits ($B_1,B_2,\cdots,B_m, C_1, C_2, \cdots,
C_n$) that are sent by Alice to Bobs and Charlies (because which one
of Bobs or Charlies will possess the secret state is previously
undefined in the scheme), and then resends fake qubits (denoted by
$B'_1,B'_2,\cdots,B'_m, C'_1, C'_2, \cdots, C'_n$) to Bobs and
Charlies. For keeping the quantum correlation among Bobs and
Charlies as good as possible, Eve may prepare the $m+n$ fake qubits
in the graph state
\begin{eqnarray}
|\mathcal{G}\rangle_{m+n}
  &=&\frac{1}{2}(|0_{B'_1}0_{B'_2}\cdots 0_{B'_m}0_{C'_1}0_{C'_2}\cdots 0_{C'_n}\rangle\nonumber\\
 &&+|0_{B'_1}0_{B'_2}\cdots 0_{B'_m}1_{C'_1}1_{C'_2}\cdots 1_{C'_n}\rangle\nonumber\\
&&+|1_{B'_1}1_{B'_2}\cdots 1_{B'_m}0_{C'_1}0_{C'_2}\cdots 0_{C'_n}\rangle\nonumber\\
&&   -|1_{B'_1}1_{B'_2}\cdots 1_{B'_m}1_{C'_1}1_{C'_2}\cdots 1_{C'_n}\rangle).
\end{eqnarray}
However, the quantum correlation between Alice and Bobs and Charlies
is destroyed. Alice, Bobs, and Charlies can easily detect such an
attack by performing suitable local measurements on the qubits they
own or receive. For example, they all select the measurement basis
$\{|0\rangle,|1\rangle\}$: under Eve's attack, there is no
correlation between the measurement outcomes of Alice and Bobs (or
Charlies); however, in the no-eavesdropping case, the measurement
outcome of Alice is always correlated (i.e., the same) with that of
Bobs and anti-correlated with that of Charlies. Thus, the
eavesdropping attack can always be detected by checking the quantum
correlation of the entanglement channel, due to the fact that
entanglement is monogamous \cite{61PRA052306}. For checking the
security, a subset of entanglement channels will be sacrificed. As a
matter of fact, most of entanglement-based quantum-communication
schemes need ones to utilize quantum correlations and sacrifice a
subset of entanglement channels to check the security against
eavesdroppers' interceptions.

In conclusion, we have proposed a multiparty hierarchical QIS
scheme, where the agents are divided into two grades ($G_1$ and
$G_2$) and the number of agents in both grades can be arbitrary in
principle. The agents of grade $G_1$ have a larger authority (or
power) than the ones of grade $G_2$ to recover the sender's secret
state. Except for sender's Bell-state measurement, no nonlocal
operation (multi-particle operation) is involved in our scheme, in
contrast to previous asymmetric QIS schemes based on the idea of
quantum error-correcting codes where multi-particle collective
operations are required
\cite{61PRA042311,64PRA042311,71PRA012328,81PRA052333}. The proposed
scheme have also been shown to be secure against eavesdropping. Our
scheme may be considered as a complementarity to conventional
(symmetric) QIS schemes without using quantum error-correcting
codes. In addition, the hierarchical QIS may be very interesting
with respect to the reliability of participants in quantum
communication and the access controlling in architecture of quantum
computer \cite{0801.1544,0906.2297,283OC1196}. The quantum channel
in our scheme is a graph state, a very important quantum resource
for quantum-information science \cite{86PRL5188,0602096,364PLA7}.
Recently, other types of QIS schemes with graph states have also
been presented \cite{78PRA042309,78PRA062333}. These schemes,
however, are very different from ours. What discussed in
Ref.~\cite{78PRA042309} are ($N$, $N$)-threshold and ($3$,
$5$)-threshold (symmetric) QIS protocols based on quantum
error-correcting codes. Ref.~\cite{78PRA062333} is focused on
two-party QIS with four- or five-qubit graph states. The key points
for physical realization of the presented QIS scheme are preparation
of graph state $|\mathcal{G}\rangle_{1+m+n}$ of Eq.~(\ref{QC}) and
Bell-state measurement. Bell-state measurement is well within state
of the art for both photon- and matter-qubits
\cite{4NPh376,443N838}. The graph state
$|\mathcal{G}\rangle_{1+m+n}$ can be efficiently generated through
realistic linear optics with the idea of Ref.~\cite{97PRL143601}.
$|\mathcal{G}\rangle_{1+2+3}$ has already been experimentally
realized \cite{3NP91}.

\section*{Acknowledgements}
This work was supported by National Natural Science Foundation of
China (Grant No. 11004050), Scientific Research Fund of Hunan
Provincial Education Department of China (Grant Nos. 09A013 and
10B013), Science and Technology Research Foundation of Hunan
Province of China (Grant No. 2010FJ4120), Excellent Talents Program
of Hengyang Normal University of China (Grant No. 2010YCJH01), and
Science Foundation of Hengyang Normal University of China (Grant No.
Grant No. 10B69). X. W. Wang thanks Chuan-Jia Shan and Xiao-Ming Xiu
for useful discussions.

\end{document}